\documentclass[11pt,preprint]{aastex}

\newcommand{\Lsun}{$L_\odot$}
\newcommand{\Msun}{$M_\odot$}

\newcommand{\ia}{\'\i}

\shorttitle{Far Infrared emission in NGC 253}
\shortauthors{Melo et al.}

\begin{document} 
\title{The spatial distribution of the far-infrared emission in NGC 253}
\author{V. P. Melo, A. M. P\'erez Garc{\ia}a, J. A. Acosta--Pulido, 
 C. Mu\~{n}oz--Tu\~{n}on \\ \& J. M. Rodr{\ia}guez--Espinosa}
\affil{Instituto de Astrof{\ia}sica de Canarias}
\affil{La Laguna, Spain}
\email{vmelo@ll.iac.es, apg@ll.iac.es, jap@ll.iac.es, \\ cmt@ll.iac.es,
jre@ll.iac.es}

\begin{abstract}

We study the far-infrared emission properties of the nearby starburst galaxy
NGC~253 based on {\it IRAS} maps and an ISOPHOT map at 180 $\mu$m.
Based on the analysis of the light profiles, 
we have been able to identify three main structural components: 
an unresolved nuclear component, an exponential disk, and a kiloparsec scale bar. 
In addition, we also found a ring structure at the end of the bar
that is particularly conspicuous at 12\micron. 
The Spectral Energy Distribution (SED) of each morphological component
has been modeled as thermal dust emission at different temperatures. 
The unresolved nuclear component is dominated by cold dust 
emission (T$_n\simeq 50$~K), whereas the disk emission is dominated by 
very cold dust (T$_{vc}^d\simeq 16$~K) plus a contribution 
from cold dust (T$_c^{d}\simeq 55$~K). 
The bar emission corresponds mainly to cold dust (T$_c^{b}\simeq 23$~K) plus a 
warm component (T$_w^{b}\simeq 148$~K).
We detect an extension of the disk emission due to very cold dust, 
which contributes a large fraction (94\%) of the total dust mass of the galaxy.
The estimated total dust mass  is $8.2 \times 10^{7}$ \Msun.

\end{abstract}


\section{Introduction}
\label{sec:Intro}

Much of our knowledge about the infrared (IR) emission from galaxies 
comes from their total emission fluxes. Models of the 
spectral energy distribution (SED),
from UV to radio wavelengths have been successful in reproducing the
observations in a variety of galaxies, from normal to starburst types
\citep{Devriendt99, Silva98, Efstathiou00}. 
Nevertheless, there are still some controversial aspects which need to be 
addressed, such as:
\begin{itemize}
\item[] Whether  heating of dust by ultraviolet photons from young stars 
is the dominant source of the far-IR (FIR) emission \citep{Devereux94}, or 
if there is a noticeable contribution from non-ionizing photons from bulge and 
disk stars \citep{Walterbos87, Xu96, Mayya97}.
\item[] The inner gas-to-warm dust mass ratios in external galaxies result in
about 5--6 times the local Galactic value
\citep{Devereux90},  implying a deficit of cold dust.
This can be attributed to a lack of data beyond 100 $\mu$m, yielding
both an  overestimate of the dust temperature and an 
underestimate of cold dust masses hidden by other, more energetically 
important, components.
\item[] The extension of dust disks as compared to stellar disks
observed in the optical bands. 
\citet{Engargiola91} and \citet{EngargHarp92} found that 
the scalelength of stellar disk is similar or larger than the observed 
at 160\micron\ continuum.  However recent work seems to support 
dust disks larger than the stellar ones
by factors from 1.4 up to 2 \citep{Davies97,Alton98,Xilouris99}. 
\item[] What is the opacity of the disk in spiral galaxies? \citet{Disney89}
suggested that the spiral galaxies could be optically thick. However, more 
recent works indicate only moderately opaque disks \citep{Xilouris99}. 
This feature has relevant cosmological implications, giving different degrees of
extinction to be accounted for in the cosmic background. 
\end{itemize}
 
Adding spatial resolution to the study of the far-IR emission from galaxies
certainly helps resolving most of these controversial points. 
This has  been possible only in a limited number of nearby galaxies thanks to 
the instruments on board the NASA--Kuiper Airborne Observatory (KAO) and  
to the improved resolution ($\sim1$~arcmin) of maps obtained 
with the {\it Infrared Astronomical Satellite} ({\it IRAS\/}) 
and processed by the HiRes algorithm 
\citep{Aumann90,Rice93}. 
\citet{Engargiola91} has studied the FIR continuum distributions of
NGC~6946 from 60 to 200~$\mu$m, using {\it IRAS} and NASA--Kuiper Airborne
Observatory data, showing that the light distribution reveals discrete
sources coinciding with giant H~{\scshape ii}
regions superimposed on an unresolved exponential dust emission background. 
The brightness profiles were separated into a
nuclear starburst plus an exponential disk. Both the nuclear and the disk SEDs
were described as modified blackbody sources at $T \sim$ 40~K and  $T \sim$ 15
K, respectively. 
Far IR maps of the quiescent spiral galaxies M~31  and 
NGC~4565  have been studied by  \citet{Xu96}
and \citet{EngargHarp92}, respectively. 
In both cases they could distinguish warm dust
tracing the spiral arms superimposed  on a diffuse cool dust emission
corresponding to the disk. They also found that old stars contribute
an important fraction of the heating of the dust. 
 
More recently, with the launch of {\it ISO}
with its on-board instrument ISOPHOT, maps of similar or better
resolution have been obtained up to 200~\micron, allowing more detailed 
studies 
of various physical quantities related to the FIR emission of galaxies. Questions  
concerning the dust  and tem\-pe\-ra\-tu\-re distributions, the structural components 
followed by the dust distribution, the existence of very cold dust emission, 
and the physical extent of the disks can all be addressed.  
\citet{Haas99} have mapped the FIR emission
of M~31 at 175~\micron\ using ISOPHOT data, showing the importance of the
contribution of cold dust to the total mass of the galaxy.

Relevant advances also come from the theoretical front. Most current 
models are able to reproduce the main features of the far IR emission from
different types of galaxies consistently with other wavelength ranges 
dominated by the stellar radiation. For instance, \citet{Devriendt99} 
developed  successful models which take into account the evolution of the
stellar population as well as extinction and emission of dust in a self
consistent way. On the other hand, 
\citet{Bianchi00} constructed bidimensional models to predict the 
dust-reprocessed FIR output in spiral galaxies corresponding to different
stellar and dust geometries. 
These models could explain the observed far IR spectral energy distribution 
from spiral galaxies only when dust disks are optically thick. However they  
failed to reproduce the large far IR scale lengths suggested by 
recent far IR observations \citep{Alton98}.

Because of its proximity, NGC~253  is a good candidate for detailed studies 
at any wavelength range. 
Its angular size in the visible range is 27.5$'$ $\times$ 6.8$'$. 
It is classified as a spiral galaxy of type SAB(s)c and
is seen almost edge--on.
Throughout this paper, we adopt a distance of
$D = 3.4$ Mpc \citep{Sandage75}, implying a scale of 16.5 pc/arcsec.
NGC~253  is considered as a prototype nuclear starburst
galaxy, exhibiting strong IR emission 
(L$_{IR}\sim 2\times 10^{10} {\rm L}_\odot$),
being an ideal target for IR studies.
In this paper, we analyze maps of NGC~253 from 12~\micron\  to 
180~\micron. We are able to separate the far IR emission into its morphological
components. The SED of each component was been modeled in terms of blackbodies
emission.

\section{Mid and Far IR maps.
%\footnote{Based on observations with ISO, an ESA project with 
%instruments funded by ESA Member States (especially the PI countries: 
%France, Germany, the Netherland and 
%the United Kingdom) and with the participation of ISAS and NASA.}
}
\label{sec:data}

\begin{figure}
\plotone{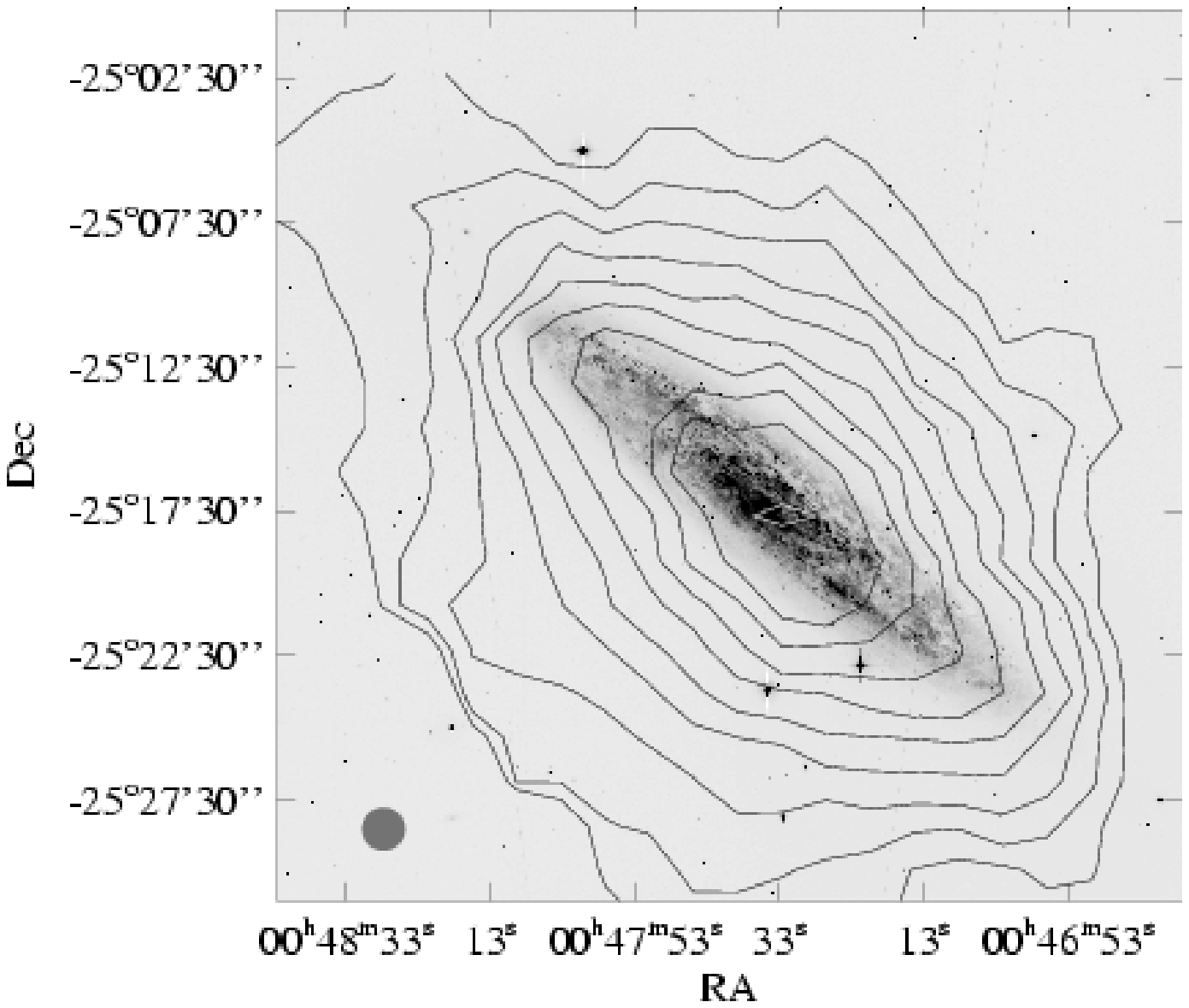}
\caption{Contours of the ISOPHOT  180~$\mu$m map overplotted on a $B$ band
image (Taken from the European Southern Observatory database). The instrumental
resolution is shown at the bottom left corner. Contours represent luminosity in
logaritmic scale and range from 3$\sigma$ to the maximum value. Contours are 
separated by a factor 1.8 from 1 to 230 MJy/sr.}
\label{fi:contornos}
\end{figure}

The ISOPHOT \citep{Lemke96} data at 180 $\mu$m were retrieved 
from the ISO\footnote{Based on observations with ISO, an ESA project with 
instruments funded by ESA Member States (especially the PI countries: 
France, Germany, the Netherland and 
the United Kingdom) and with the participation of ISAS and NASA.} 
Data Archive. 
The map was performed in undersampled raster mode with a step of
180~arcsec,
using the C200 camera, which consists of a 2 $\times$ 2 pixel array (the pixel
size is 89~arcsec). 
The data were reduced using PIA v9.1 \citep{Gabriel97}, and  
the following corrections were applied: 
ramp li\-neari\-za\-tion, deglitching at the ramp and signal levels, and
dark signal subtraction. Only the last 60\% of the signal at each raster 
position was  retained in order to minimize transient effects. 
The flux calibration
was done using the internal calibration lamps, although 
the final calibration was determined by comparing the sky background 
measurement in the  180 $\mu$m ISOPHOT map with background measurements from 
{\it COBE\/}/DIRBE maps. A scaling factor of 1.48 has to be applied to the 
ISOPHOT data in order to match both numbers. 
Two pixels of the C200 array were saturated when crossing the nucleus of 
the galaxy. The  values at these map positions were obtained by interpolation from
the neighbourings . This fact may produce a broadening of the light profile 
at the center of the galaxy, and probably underestimate the nuclear contribution.
The map reconstruction was performed within PIA using the trigrid algorithm. 
We estimate the actual resolution of this map to be  $\sim 180$~arcsec, due
to the poor raster sampling. 
The isocontours of the 180~\micron\ image overlaid on a $B$ band 
(MPG/ESO 2.2 m + WFI optical image\footnote{Obtained with the WFI (full field) 
at the MPG/ESO 2.2 m telescope and retrieved from the ESO archive.})
are presented in Fig.~\ref{fi:contornos}. The main features present in 
the optical image can also be recognized in the 180~\micron\ map. 
However, the morphology at 180\micron\ is slightly different than 
that observed in the other IR bands. The emission along the minor axis extends 
up to 10~Kpc from the nucleus. This extended emission can be associated with a 
cold dust halo. Addittional evidence was found in our analysis of the strip maps
perpendicular to the disk \citep{Perezetal01} The same conclussion has been 
reached by \citet{Radovich01}. A similar morphology has been reported from 
ROSAT X-ray observations by \citet{Pietsch00}.

\notetoeditor{Figure 1 here.}

\begin{figure}
\epsscale{0.88}
\plotone{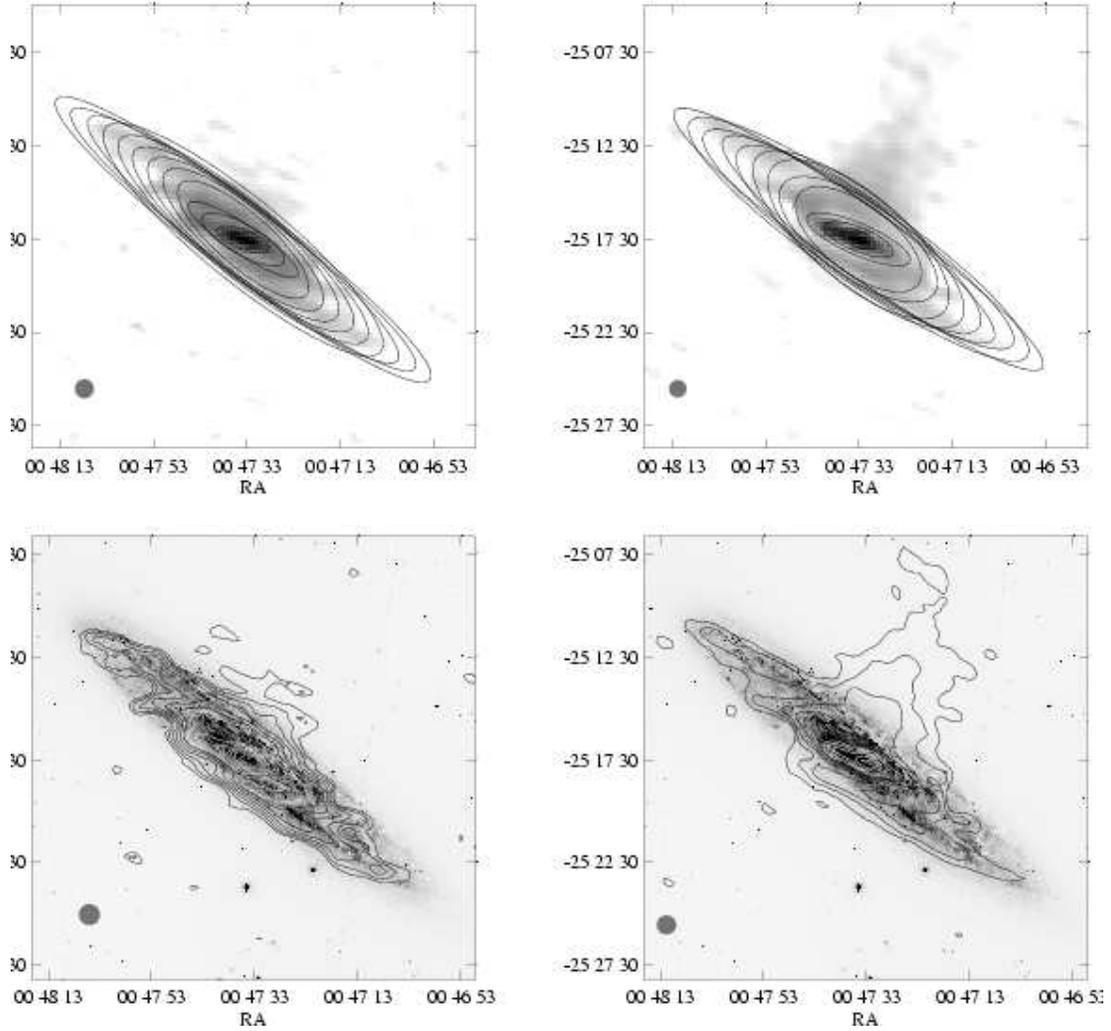}
%\epsscale{0.38}
%\plotone{figure2b.ps}
\caption{Left panel: Down: Contours of the {\it IRAS} image at 12~$\mu$m 
overplotted on a $B$ band image (Taken from the European Southern 
Observatory database).Contours represent luminosity in
logaritmic scale and range from 5$\sigma$ to the maximum value. Contours are 
separated by a factor 1.8 from 1.3 to 340 MJy/sr. Up: IRAS image at 12$\mu$m
with overlaid ellipse fits.
Right panel: Up: {\it IRAS} image at 25 $\mu$m, with overlaid ellipse fits. 
Down: Contours represent 
luminosity in logaritmic scale and range from 5$\sigma$ to the maximum value. 
Contours are separated by a factor 1.8 from 1.5 to 2100 MJy/sr. The 
instrumental resolution is shown at the bottom left corner of each panel.}
\label{fi:isofotas}
\end{figure}

We retrieved {\it IRAS} maps at 12, 25, 60, and 100$\mu$m from the Infrared
Processing and Analysis Center at Caltech (IPAC). 
The data retrieved were processed using  
HiRes algorithm, which is based on the Maximum Correlation
Method \citep{Aumann90} to produce images with improved spatial resolution. 
The resolution of these IRAS maps ranges from 1~arcmin at 12 and 25~\micron\ 
to 1.7~arcmin at 100\micron.
Unfortunately, the low level emission seen in the {\it IRAS} maps cannot 
be used due to the spurious structures caused 
at 12 and 25 \micron\ (see Fig \ref{fi:isofotas}) by detector hysteresis 
after crossing the bright nucleus,
and at 60 and 100 \micron\  by reflected emission off the {\it IRAS} 
secondary spider \citep{Rice93}. 

Very recently, \citet{Radovich01b} have studied the  far IR emission 
of NGC~253 using IRAS and ISO data, as well as a
numerical radiative transfer model. Their main aim war detecting FIR emission 
above the disk that might be connected with outflows and/or supergalactic winds. 
These authors have also shown the need for an extended dust disk.

% Fig 2 : Contours+isophotes over greyscale images
\notetoeditor{Figure 2 here.}

\section{Results}

\subsection{IR luminosity profiles}
\label{sec:profiles}

\begin{figure}
\epsscale{0.8}
\plotone{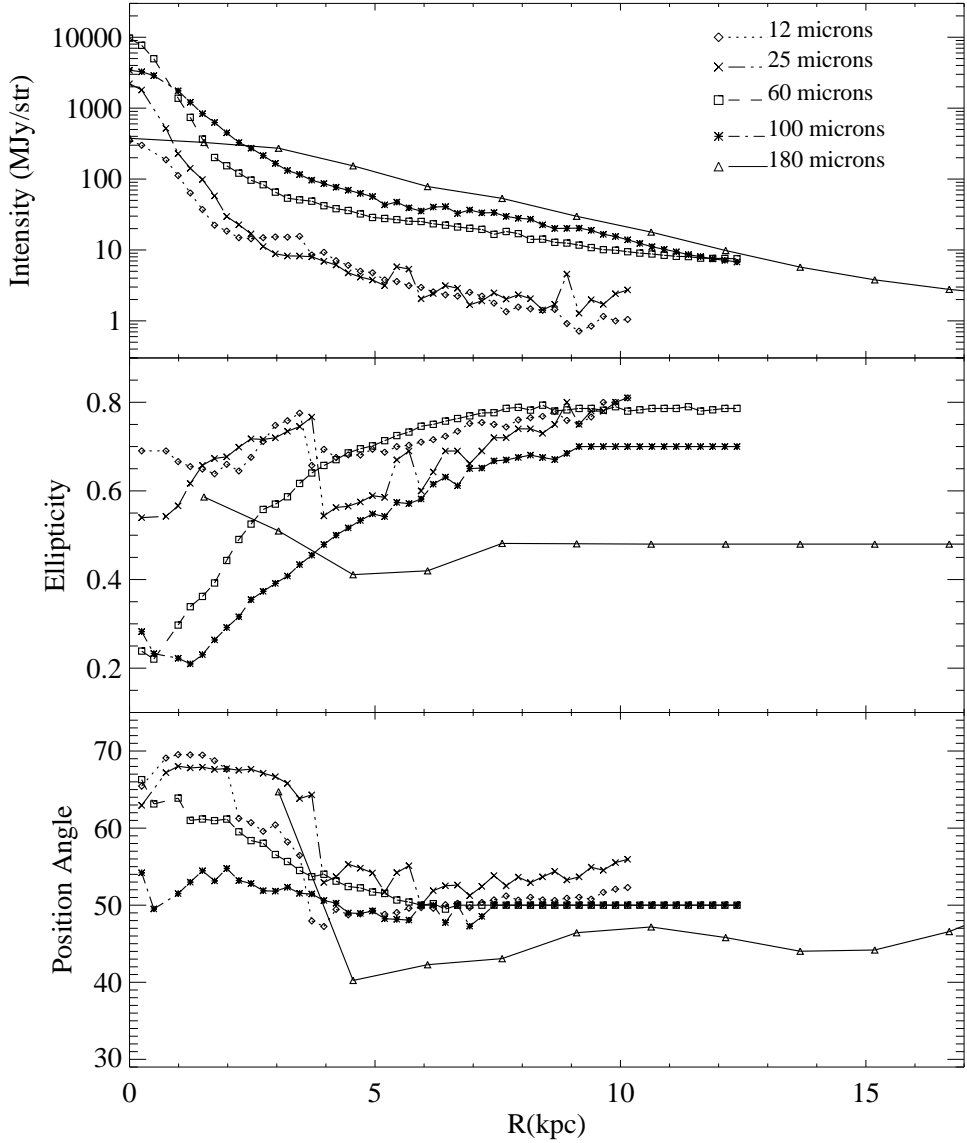}
\caption{Radial variation of the  elliptical isophote parameters  from the
different filters versus radial distance from the center. The different lines 
representation is as follows: dotted line for 12~\micron, 25 
dash-dot-dashed line for 25~\micron, dashed line for
60~\micron, dashed-dotted line for 100~\micron, and continuous line for 
180~\micron. 
From top to bottom we present the luminosity profiles, ellipticity and 
position angle.}
\label{fi:radparam}
\end{figure}

\begin{figure}
\epsscale{0.9}
\plotone{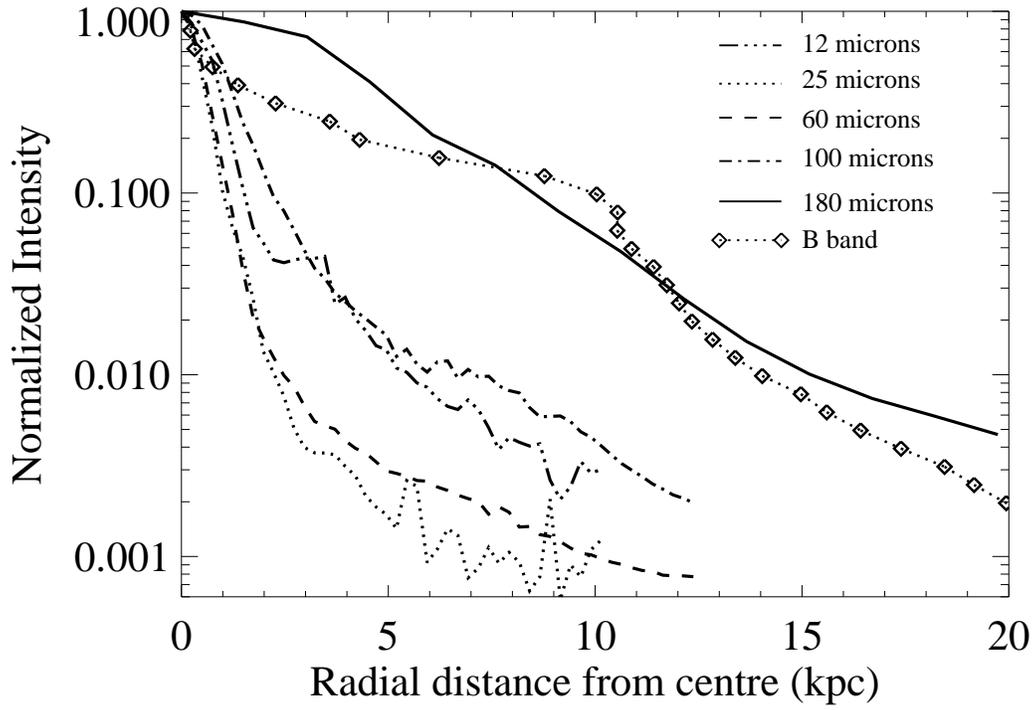}
\caption{Normalized luminosity profiles at 12 $\mu$m  (dashed-dotted line),
25 $\mu$m  (dotted line), 60 $\mu$m  (dashed line), 100$\mu$m  (dash-dot-dashed
line), and 180~$\mu$m  (continuous line). 
The luminosity profile in the $B$ band is also shown (connected diamonds). 
The B data are taken from \citet{Pence80}. Note the slow decay  of the
180~\micron\ emission, comparable to the B emission, in contrast to the 
sharp profiles in the range 12 to 100~\micron.}
\label{fi:radprof}
\end{figure}

For all maps, we obtained surface brightness profiles by azimuthally averaging 
over elliptical annuli. We fitted the isocontours of surface brightness 
with the ELLIPSE task in IRAF\footnote{IRAF is the Image Reduction
and Analysis Facility, a general purpose software system for the
reduction and analysis of scientific data. IRAF is written and
supported by the IRAF programming group at the National
Optical Astronomy Observatories (NOAO).}, 
which uses the algorithm described in \citet{Jedrzejewski87}. 
The spatial interval between two consecutive isophotes is equal to the pixel size in
each filter, but averaged over an annulus of width close to
the resolution in each filter. 
Figure \ref{fi:isofotas}  shows the family of ellipses fitted for two filters, 
12 and 25 \micron. At first glance, it can be noticed  how
the isophotal twisting delineates the presence of a bar 
and a ring at the circumnuclear region (discussed below).
The brightness profiles, ellipticities ($\epsilon$), and position angles (PA) 
in all filters as a function of radial distance from the galaxy center are 
shown in Fig. \ref{fi:radparam}. 
Fig. \ref{fi:radprof} shows the light profiles normalized to
the central peak to facilitate the comparison at the different
wavelengths. The brightness profiles show similar behavior in all  the 
IR filters, with the exception of the 180~\micron\ profile.  
Two different slopes are clearly distinguished, a steep profile 
from the central peak out to about 2--4 arcmin, and from there on 
the profile flattens sampling the disk of the galaxy (see Fig. \ref{fi:radprof}). 
The IR emission concentrates at different radii
depending on the wavelength: the sharpness of 
the luminosity profiles from 12 to 100~\micron\ reflects 
the predominance of the warm dust emission associated 
with the starburst activity at the galaxy center. In the outer parts the 
disk is the dominant structure. 
In contrast, the luminosity profile at 180~\micron\ appears flatter than 
the other profiles, following an exponential disk profile behavior at all
radii.
No central region can be distinguished in the 180\micron\ profile.
%This profile traces the distribution of the cold dust emission, which 
%decays by 1.5 dex at a distance of 15 kpc from the nucleus.  
The absence of a central peak at 180\micron\ is probably a consequence of 
the lower prominence of the nucleus at that wavelength, although the saturation
problems mentioned above likely blur the central peak. 
Remarkably, the extension and the decay of the light profile
at 180~\micron~ is comparable to that measured 
in the {\it B} band \citep{Pence80} (see Fig \ref{fi:radprof}), although
the slope beyond 15 kpc is slightly flatter at 180~\micron~than at the B band.
This fact may indicate an extended disk of very cold dust \citep{Radovich01b}.

% Fig 3 : Light profiles, ellipticity and PA
\notetoeditor{Figure 3 here.}

More structures can be distinguished in the brightness profiles. 
At a distance of $\sim 3.5$~kpc, there is a shoulder in the 12~\micron\ 
profile, which is also evident in the 25~\micron\ profile; this feature 
coincides with the end of the inner disk detected by \citet{Scoville85},
which we have also identified  as a ring structure (discussed in section 
\ref{sec:decom}). There are  other findings which support this identification: 
an excess of millimeter CO emission at
3.5~kpc along the semi--major axis, reported by \citet{Scoville85},
and more recently by \citet{Sorai00}; 
and a H$\alpha$ image shows a remarkable ring structure with radius of 
2.7~arcmin \citep{Hoopes96}.
The prominent hump observed in the B band at a distance
between 6 and 12 arcmin along the semi--major axis is marginally present 
in the 100 $\mu$m  profile. The hump at the B band has been attributed to young stars 
and H{\scshape ii} regions in the spiral arms \citep{Pence80}.  
Note that in the other {\it IRAS} filters we cannot see 
the hump, probably because of noise introduced by spurious 
structures at low flux levels.

The position angle and the ellipticity also vary considerably with 
galactocentric distance.
In particular, the 12~\micron\ light profile shows drastic changes in both 
ellipticity and position angle at radii between 2 and 3.5 kpc. 
The ellipticity increases from the center up to a maximum value of 0.8 
at a distance of 3.5 kpc, where it suddenly decreases to a value of 0.5; 
from there outwards, the ellipticity approaches 
the disk value of 0.75. Simultaneously, the position angle changes 
from 70$^\circ$ to 60$^\circ$ at 2~kpc falling to the outer disk average 
value of 52.3$^\circ$ at a distance of 3.5 kpc. 
The 25 $\mu$m profile show a similar behavior to that at 12 $\mu$m,
although the position angle remains constant at 70$^\circ$ 
up to 3.5 kpc. 
We identify these features with the presence of a bar, whose existence
has been reported previously by other authors 
\citep{Pence80,Scoville85,Forbes92}. 
Both the orientation and the extension match  well
the values reported in the near IR: \citet{Scoville85} found in 
a 2.2 \micron\ map an elongated
barlike structure with a radius $\sim 1.98.$~kpc, oriented at PA = 68$^\circ$,
within an inner disk which extends out to $\sim 2.97$~kpc;
\citet{Forbes92}  found in an $H$ band
image a bar oriented at $\sim 70^\circ$ and extending 2.48~kpc.  
The barred nature of this galaxy is obvious in the near IR \citep{Scoville85}
and also quite evident 
in the 2MASS Atlas Image Gallery mosaic of this galaxy\footnote{This image
(J,H \& Ks composite) is
available electronically at: 
http://\-www.ipac.caltech.edu/\-2mass/\-gallery/\-images\_galaxies.html.}.
Moving outwards, the position angle and ellipticity reach  
the same asymptotic limiting values for the different filters, 
within the errors.
We obtain for the ellipticity the average value 0.79 $\pm$ 0.10, which 
corresponds to a galaxy inclination of $i \simeq 78^\circ \pm 6^\circ$, 
and position angle $50 ^\circ \pm 7$,  representing
the disk. These values are in good agreement with those obtained by
\citet{Pence80} using optical data 
(PA = 51$^\circ$, $i \simeq 78.4^\circ \pm 0.3^\circ $).
As the resolution decreases---moving to longer wavelengths---details of the 
isophotes at the inner part become less representative. 
At 180 $\mu$m,  the position angle reaches  similar values in the outer parts
to those at 12, 25, 60, and 100 $\mu$m. 
However the ellipticity remains lower,  the isophotes being rounder. 
This is consistent with the detection of a cold dust halo. This halo is
not detected at other wavelengths except at 100 \micron\ which shows a
slight lower ellipticity. This is consistent with the work by \citet{Alton98b}
who fail to detect dust emission above the disc using IRAS images from edge-on
galaxies.

% Fig 4 : Normalized luminosity distributions
\notetoeditor{Figure 4 here.}

% Fig 5 : Variation of IR colors
\notetoeditor{Here Figure 5.}

\begin{figure}
\epsscale{0.8}
\plotone{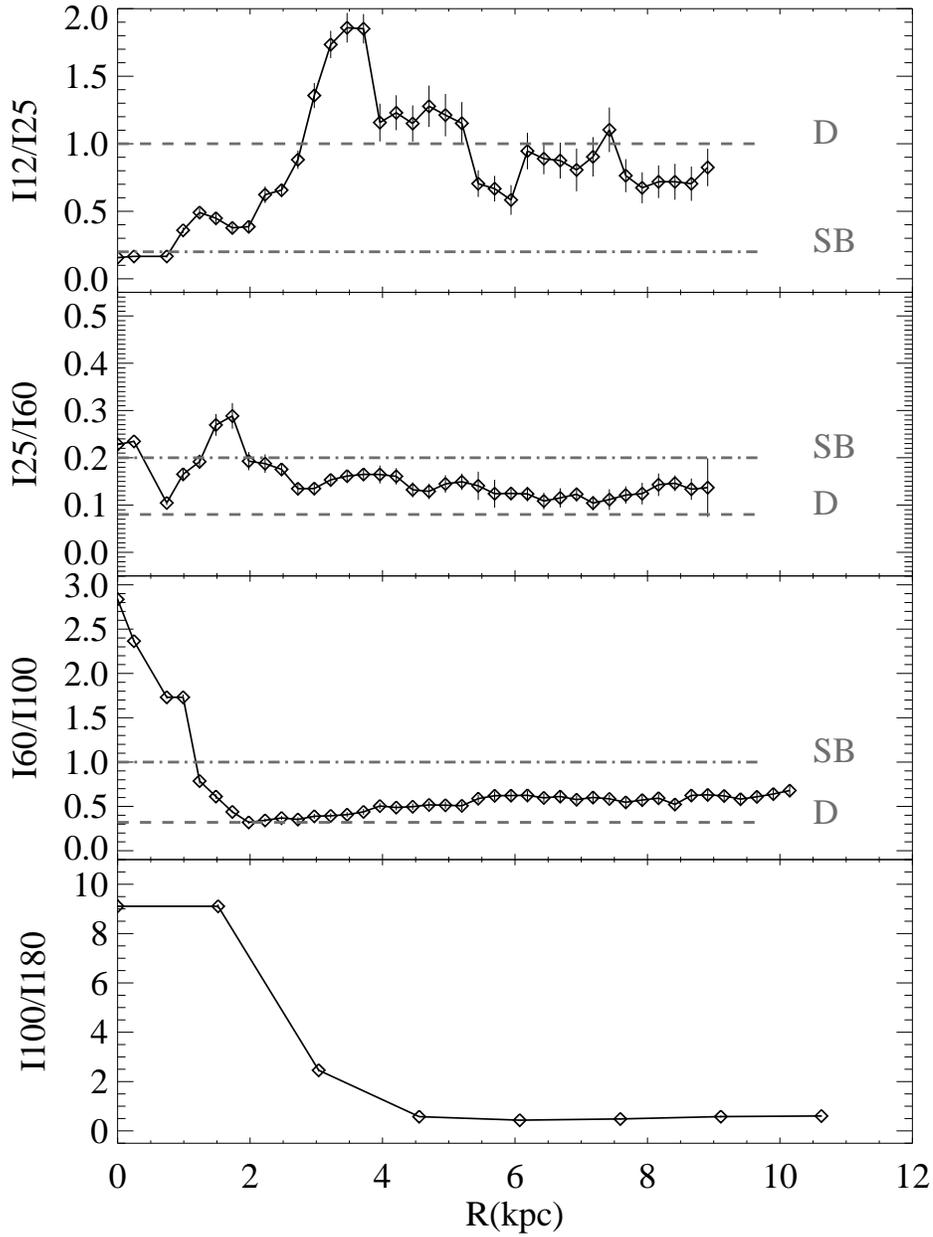}
\caption{Variation  of IR colors along the  
major axis of elliptical isophotes. The data points are smoothed 
to the worse resolution in each color combination. 
Horizontal lines are the values expected 
for starbursts (SB) and discs (D) from \citet{Rowan89}. }
\label{fi:radcolor}
\end{figure}

We have also looked at the variation of the IR colors with galactocentric
distance (see Fig. \ref{fi:radcolor}). The mid- and far-infrared 
color ratios approximately describe  the origin of the dust emission
at these wavelengths \citep{Helou00, Telesco93}. 
In a simplified scenario, the IR emission 
of a  relatively normal galaxy can be described as a composite of two 
spectral components \citep{Helou86}:  blackbody emission from 
classical grains in temperature equilibrium heated by the interstellar radiation
field dominating in the far IR;
and very small grains and PAH molecules transiently heated to high 
temperatures by the absorption of single energetic photons produced in star
forming regions.  As a result  
starbursts have warmer 
60/100~\micron\ color temperatures and cooler 12/25~\micron\ color
temperatures. The more active starbursts exhibit 
bluer 25/60~\micron\  and 60/100~\micron\ colors corresponding 
to warmer dust in high temperature equilibrium. As a guide, we have 
superposed for all colors  the typical values of two model components---the 
disk (D) and the starburst (SB). These models were used to synthesize the 
observed far-infrared spectra of many IRAS galaxies by  \citet{Rowan89}. 
The first thing to notice is that in the outer regions all colors are 
typical of disk emission. The 12/25~\micron\  color index shows
a strong peak,  coincident with well identified morphological features
(described in next section).
In the innermost regions, this color matches
that of typical starburst galaxies. 
Around 3--4 kpc there is a strong color increase  
slightly beyond the end of the bar  at the location of the ring
detected in our profile separation (as described below). 
The origin of this color enhancement can be related to an intense PAH
emission associated with the ring of molecular gas mentioned above
\citep{Scoville85,Sorai00}.
The  25/60~\micron\ color index shows a smoother trend. 
An interesting feature appears around 2~kpc, coincident with the head of the
barlike structure, that might be another
indication that star formation is taking place there.
The  60/100~\micron\ and 100/180~\micron\ color indexes have 
an absolute maximum at the center of the galaxy, 
as corresponds to the central starburst activity.  
The  60/100~\micron\ color index shows a small increase coincident 
with the beginning of the spiral arm. 
This is interpreted as an increase in dust 
temperature due to star formation in large HII regions, as
observed by \citet{Engargiola91} in NGC 6946.

\subsection{Surface brightness profile decomposition}
\label{sec:decom}

\begin{figure}
\epsscale{0.8}
\plotone{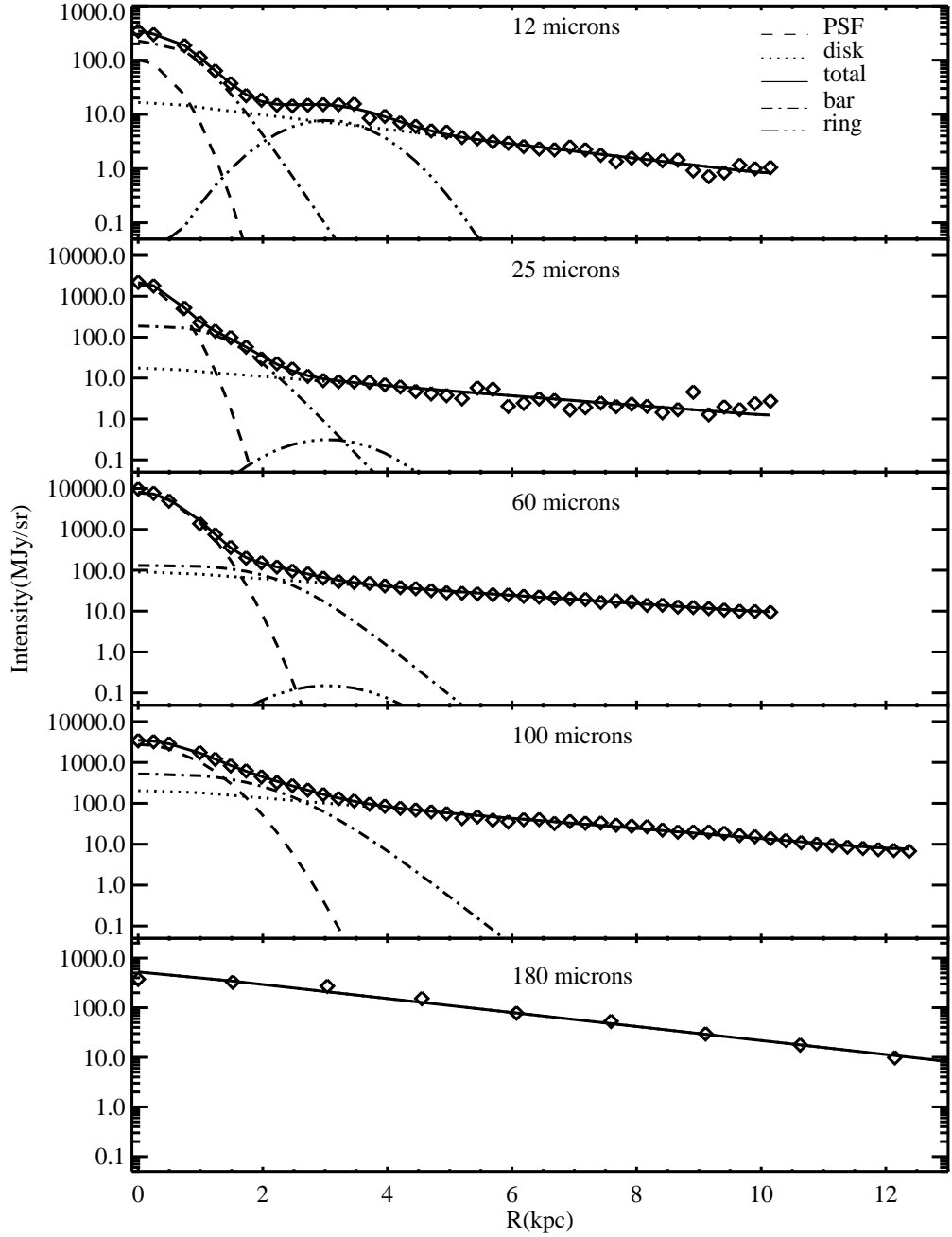}
\caption{Surface brightness profile decomposition for all bands: 
12, 25, 60, 100, and 180 $\mu$m profiles (from top to bottom).
The unresolved nuclear component (dashed line) dominates the central emission
except at 180 $\mu$m (see text). The bar (dash-dot-dashed line) 
is clearly distinguished up to
100~\micron. The disk (dotted line) always appears at the outermost isophotes.  
The ring (long dash-dotted line) is clearly seen only at 12 and 25~\micron.}
\label{fi:perf}
\end{figure}

The best fit structural components to the radial light profiles at 12, 25, 60,
100 \& 180$\mu$m are shown in F. \ref{fi:perf}.
Initially, we  tried a morphological separation of the
profiles into two components:  an unresolved source 
corresponding to the PSF (for the {\it IRAS}
 maps we use gaussian functions and 
for the ISOPHOT map a model PSF) and an exponential disk that
mainly accounts for the external regions of the galaxy. The resolution of the
180~\micron\ map is not sharp enough to separate any other components than the 
exponential disk. 
A bulge component was not included because it cannot be sampled with the 
limited resolution of our maps. \citet{Forbes92} found an effective radius
(10$''$)
for the bulge  which is well below our resolution limit, so
 its contribution will be implicitly included in the 
unresolved component. 
The  analytic profiles were convolved using the corresponding PSF 
function for each filter before performing the fit. 
A third component was introduced to improve the matching of the models with
the observed profiles following the indications of the presence of a
bar discussed in Section \ref{sec:profiles}.
For this latter component we have taken a flat bar profile
\citep{Prieto97}: 
$$  I_{\rm bar}(r)=\frac{I_{\rm bar}}{1 + 
  \exp\left({\frac{r-r_{\rm bar}}{l_{\rm bar}}}\right)}, $$ 
where $I_{\rm bar}$  is the amplitude, $r_{\rm bar}$ is the length and
$l_{\rm bar}$ is the downward gradient.   

%\documentclass[10pt,preprint]{aastex}
%\begin{document}

\begin{deluxetable}{llllllllll}
\tabletypesize{\scriptsize}
\tablecaption{Structural parameters\label{tbl-1}}
\tablewidth{0pt}
\tablehead{
\colhead{$\lambda$} & 
\colhead{I$_{PSF}$}  & 
\colhead{I$_{disc}$}   & 
\colhead{r$_{disc}$} & 
\colhead{I$_{bar}$}  &
\colhead{r$_{bar}$} & 
\colhead{l$_{bar}$} & 
\colhead{I$_{ring}$}  & 
\colhead{r$_{ring}$} & 
\colhead{l$_{ring}$} 
\\ 
\colhead{$\mu m$} & \colhead{MJy/sr} & \colhead{MJy/sr} & \colhead{kpc}
 & \colhead{MJy/sr} & \colhead{kpc} & \colhead{kpc} & \colhead{MJy/sr} 
 & \colhead{kpc} & \colhead{kpc}}
\startdata
12 & 100$\pm$20 &   18$\pm$2 &       3.2$\pm$0.3 &   280$\pm$40 &
0.73$\pm$0.08 &     0.26$\pm$0.03 &    8.5$\pm$0.9 &   3.0$\pm$0.3  &
0.69$\pm$0.11  \\

25 &   1950$\pm$200 & 19$\pm$2 & 3.7$\pm$0.4&  
      190$\pm$20  &    1.34$\pm$0.14 &   0.28$\pm$0.03 & 
      0.3$\pm$0.4 &    3.0$\pm$0.3  &   0.69$\pm$0.11 \\

60 &   7880$\pm$790    &   99$\pm$10 &  4.3$\pm$0.4 &
  130$\pm$20   &    2.1$\pm$0.2 &  0.35$\pm$0.12 &
  0.2$\pm$3.9  &   3.0$\pm$0.3  &   0.69$\pm$0.11 \\
         
100&  2760$\pm$280   &    230$\pm$20  & 3.5$\pm$0.4 &
 550$\pm$70 &   1.9$\pm$0.2 &  0.37$\pm$0.04 &
 4.8e-4$\pm$2.4e-4  &  3.0$\pm$0.3  &   0.69$\pm$0.11  \\

180&   ...    &   550$\pm$170   & 3.1$\pm$0.9 & 
      ... &   ... &  ...  &  
      ... &  ...  &	  ... \\

\enddata

\end{deluxetable}

%\end{document}

In addition, the 12 $\mu$m luminosity profile shows a conspicuous maximum 
around 3.5~arcmin, at the position where a strong decrease 
in the ellipticity and a noticeable variation in the position angle are found. 
We identify this feature with a ring at the end of the bar,
which also coincides with the maximum extent of the inner disk
identified by \citet{Scoville85}. This
structure can be related to the molecular
and H$\alpha$ rings detected by\citep{Sorai00} and \citep{Hoopes96}.
A ring structure can be described by a Gaussian profile 
\citep{Buta96}, as follows:
$$  I_{\rm ring}(r) = I_{\rm ring}\exp\left\{ -\frac{1}{2} 
\left(\frac{r-r_{\rm ring}}{l_{\rm ring}}\right)^2\right\}, $$
where $I_{\rm ring}$  is the amplitude, $r_{\rm ring}$ is the center and
$l_{\rm ring}$ is the width.   

The ring can still be 
recognized in the intensity profile at 25~\micron, but it disappears at 60, 
100 and 180~\micron. 
Nevertheless, we have introduced this morphological component in the 
decomposition of the 25 and 60\micron\ profile, although the 
radial distance ($r_{\rm ring}$)  and the width ($l_{\rm ring}$) are fixed 
to that at 12\micron, in order to obtain a meaningful fit. 
We have checked that introducing this component at the other wavelengths 
(60 and 100~\micron) the fit results do not change, being compatible with 
a very low amplitude for it. The uncertainty of the amplitude of the ring 
at 60\micron\ is 
compatible with non--detection (see Table \ref{tbl-1}) the seeing ring.

The resulting parameters for the different morphological components and filters 
are given in Table \ref{tbl-1}.
During the fitting process all parameters 
were allowed to vary freely, except for the radial distance and  width of the
ring, as mentioned above.
The uncertainties in the parameter values were computed using a bootstrapping 
technique. This procedure is based on a Monte Carlo 
simulation: new light profiles
are obtained by perturbation of each measured profile value, using 
a normal distribution of the same width as the error of that point. 
New fit parameters are determined 
for each simulated profile (usually 50 simulations) and the uncertainty of each
parameter is taken as the standard deviation of the resulting values. 

\notetoeditor{Here Figure 6.}

The disk scale lengths obtained in all the examined maps are very similar and
have a mean value of 3.63~kpc. This  is very close to the value
of 3.32 kpc obtained in the $B$ band by \citet{Pence80}. \citet{Puche91} 
found in H~{\sc i} observations a disk scale length of 3.41~kpc. 
However, \citet{Forbes92} obtained in the $H$ band a disk scale length of
2.65~kpc along the semi-major axis. 
The smaller  scale length observed in the $H$ band, if confirmed 
at $J$ and $K$,  could
be attributed either to high extinction at the inner parts or 
 to the presence of very high temperature small dust grains emitting in the
near IR. An important result is that both the stars scale length
\citep{Pence80} and the dust scale length are similar,
which is not in agreement to the results finding of \citet{Radovich01b} based
on radiative transfer models of this galaxy.

\subsection{Spectral energy distributions of the structural components}
\label{sec:sed}

\begin{figure}
\epsscale{0.9}
\plotone{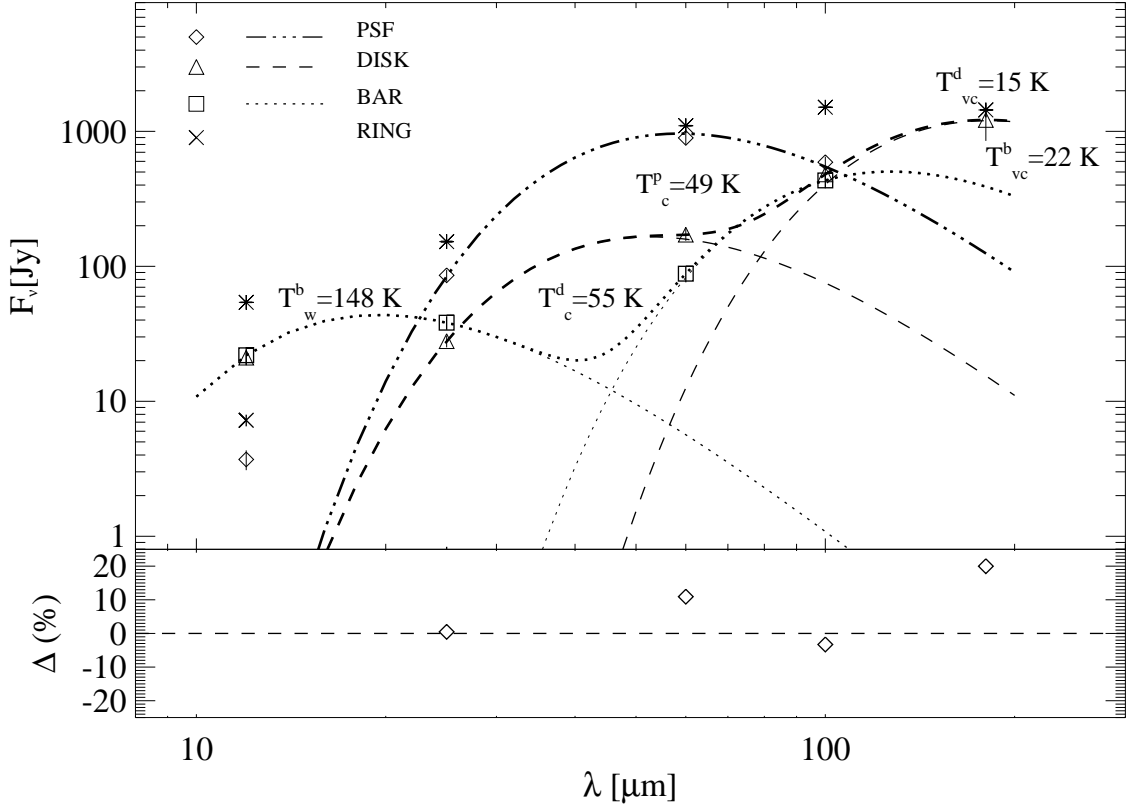}
\caption{Spectral energy distribution of the morphological components.
The SED of the disk (dashed line) is modeled as the emission of two 
blackbodies  with
T $\sim 55$~K and T $\sim$ 15 K. The SED of the nuclear component 
(dashed-dotted line) corresponds to a blackbody of T $\sim 50$~K. 
The SED of the bar 
(dotted line) is decomposed into two components peaking at T $\sim 150$~K and 
T $\sim$ 22 K. In the bottom panel we present the residuals between the
sum of models and the total emission at each wavelength.
There is a large residual at 12~\micron\ because there the emission is 
mainly attributed to PAH and very small grains which are in thermal equilibrium.}
\label{fi:bbcomp}
\end{figure}

The IR spectral energy distribution (SED) of the morphological 
components, namely, the unresolved nucleus, the disk,  and the bar found in the
previous section, are displayed in Figure \ref{fi:bbcomp}.
For each filter and each morphological component we represent the emission
integrated over the galaxy. The uncertainty of the emission from any component
is determined from the uncertainty of the amplitude.   
The global emission of each morphological component has been modeled 
as the combination of the emission from emissivity--weighted 
blackbody components \citep{Hildebrand83}. An emissivity index of 2 
has been adopted. The best fitting blackbody temperatures are compiled in 
Table \ref{tbl-2}. The uncertainties in the temperature and  
scaling factor of each blackbody were computed using the 
bootstrapping technique, similar to what was done for the profile decomposition. 

\notetoeditor{Here Figure 7.}

The SED of the unresolved nuclear component can be modeled as 
a cold ($T\simeq 49$~K) component. 
The cold component has a temperature typical of dust heated in 
star forming regions. This component is present in nearly all 
classes of galaxies, including normal, starburts, and even active galaxies 
\citep{Knapp96,Chini92,Klaas97,Perez01}. 
There is an excess at 12\micron\ which is likely due to a combination of 
PAH and very small grains emission. In fact, PAH emission
has been detected at the central regions of NGC~253 
\citep{Perezetal01,Dudley99,Keto99}

The SED of the disk component is explained by a combination of 
a cold ($T\simeq$ 55~K) and a very cold ($T\simeq$ 15~K) component. 
The origin of the warm component is again due to dust heated in H~{\sc ii}
regions and OB associations populating the arms of the disk. 
The temperature of the very cold
component is typical of dust heated by dilute interstellar
radiation. This very cold dust has been observed in normal galaxies 
\citep{Walterbos96,Walterbos87,Cox86} and
in the disks of starbursts and active galaxies \citep{Radovich99,Perez01}. 

The SED of the bar component can  be modeled as the sum of two modified
blackbodies,  a warm one at $T = 148$~K and a cold one at $T = 22$~K. 
The origin of the warm component is related to regions of very intense 
star formation \citet{Klaas97,Lutz96}. Instead,   
the temperature of the cold component indicates an origin at a quiescent 
interstellar gas region, where  star formation activity is low.

Most of the  emission from the ring is concentrated in the mid-IR range, 
from 12  to 25~\micron; beyond this wavelength the emission cannot
be reliably determined. The characteristics of the ring emission seems
to indicate a dominant contribution of PAH and very small grains, as 
already mentioned in Section 3.1, indicative of photodissociation regions.

\section{Derived physical properties}

\subsection{IR luminosities}
\label{sec:mass}

We have computed the IR luminosities between 1 and 1000 \micron\ 
for each morphological component (except for the ring because of the  
reasons mentioned in the previous section). The  luminosity for each 
morphological component can be easily computed after the blackbody model is
found.  
The resulting values are compiled in Table \ref{tbl-2}, the total values
are obtained as the sum over all structural components.

%\documentclass[12pt,preprint]{aastex}
%\begin{document}

\begin{deluxetable}{llllll}
\tabletypesize{\scriptsize}
\tablecaption{Temperatures, infrared luminositites, dust masses and star
formation rates derived for the different components, as well as 
the global values for the galaxy. \label{tbl-2}}
\tablewidth{0pt}
\tablehead{
\colhead{Component} & \colhead{T(K)} &  \colhead{F$_{IR}$}  &
\colhead{L$_{IR}$}  & \colhead{Dust mass} & 
\colhead{SFR} \\
\colhead{} & \colhead{(K)} & \colhead{[10$^{-8}$ erg~s$^{-1}$~cm$^{-2}$]} &
\colhead{[10$^9$ L$_\odot$]} & \colhead{[10$^6$ M$_\odot$]} & 
\colhead{(M$_\odot$/yr)} }
\startdata
PSF   & 49.4$\pm$0.9 & $5.77~\pm~0.03$ & 
        $20.9~\pm~1.5$  & 
	$0.21~\pm~0.02$  & 3.5$\pm$0.3 \\
	
DISK  & 55$\pm$15  & $1.097~\pm~0.011$ & 
        $4.0~\pm~1.3$  &
        $0.02~\pm~0.03$ & 0.7$\pm$0.2 \\
	
	& 15.9$\pm$2.1  &$2.4~\pm~1.6$ &
	$8~\pm~3$ & 
	$77~\pm~58$ &  \\
	
BAR   & 148$\pm$74   & $0.8~\pm~0.5$ & 
        $3~\pm~2$ &
        $<<$~1   & \\
      
      & 23$\pm$12   & $1.4~\pm~0.9$ & 
        $5~\pm~3$ &
        $5~\pm~14$  & 0.8$\pm$0.6 \\
      
TOTAL GALAXY & & $11.4~\pm~1.5$ & $41.1~\pm~1.6$  & 
$82~\pm~7$ & 5$\pm$1 \\
\enddata

%% Text for table notes should follow after the \enddata but before
%% the \end{deluxetable}. Make sure there is at least one \tablenotemark
%% in the table for each \tablenotetext.

%\tablenotetext{a}{Sample footnote for table~\ref{tbl-1} that was generated
%with the deluxetable environment}
%\tablenotetext{b}{Another sample footnote for table~\ref{tbl-1}}

%\tablecomments{Occasionally, authors wish to append a short
%paragraph of explanatory notes that pertain to the entire table, but
%which are different than the caption.  Such notes should be placed in
%a {\tt tablecomments} command like this.}

\end{deluxetable}

%\end{document}

The nuclear unresolved component is the most important
contributor to the total IR luminosity, accounting for about half of the total
emission.
The disk accounts for $\sim 30$\% of the total, of which 2/3 corresponds to 
the very cold component. The remaining $\sim 20$\% comes
from the bar structure, in which the cold component dominates. 
The contribution of the ring to the total luminosity is very small.

\subsection{Star formation rate}

Star formation rates (SFRs) were calculated following \citet{Kennicutt98} 
for both the warm and cold components using the following expression:

$$ {\rm SFR} = 1.7 \times 10^{-10} L_{\rm IR}/L_\odot \hspace{1cm} 
[M_\odot\  {\rm yr}^{-1}]. $$

This relationship applies only to relatively young starbursts,  
which implies that it only produces meaningful results when
$T_{dust} \gtrsim 25$~K.
The resulting values are compiled in Table \ref{tbl-2}. 
As expected, the maximum star formation activity is taking place in the
nuclear region at a moderate rate $\sim 3.5\ M_{\odot}/$yr. Furthermore, 
star formation activity at a lower rate is taking place in both the disk and
the bar. 
The total SFR in the galaxy is 5.0~\Msun/year.
This value is 20\% lower than that found by \citet{Radovich01b} who give a 
total
SFR value of 6.1~\Msun/yr (after correcting their value for the different
distance used herein). 
Notice that \citet{Radovich01b} include the 20~K dust emission to complete the
global SFR.
We therefore consider the value here provided as a rather more precise
value.
If we compare this value (4.97~\Msun/year) with 
that obtained from the H${\alpha}$ emission, SFR = 0.6192 \Msun/yr by
\citet{Hoopes96},
we estimate that the H${\alpha}$ emission must be obscured by a factor 
of $\sim 8$, in good agreement with the value estimated 
by \citet{Hoopes96}.

\subsection{Dust masses}

Dust masses were estimated following \citet{Hildebrand77} for the emissivity 
index ($\beta=2$) used here, assuming dust grain properties as in \citet{Hildebrand83}:

$$ M_d = 3.6 \times 10^{-5} (T_{K}/40)^{-6} L_{\rm IR}/L_\odot 
\hspace{1cm} [M_\odot] $$

\begin{figure}
\epsscale{0.9}
\plotone{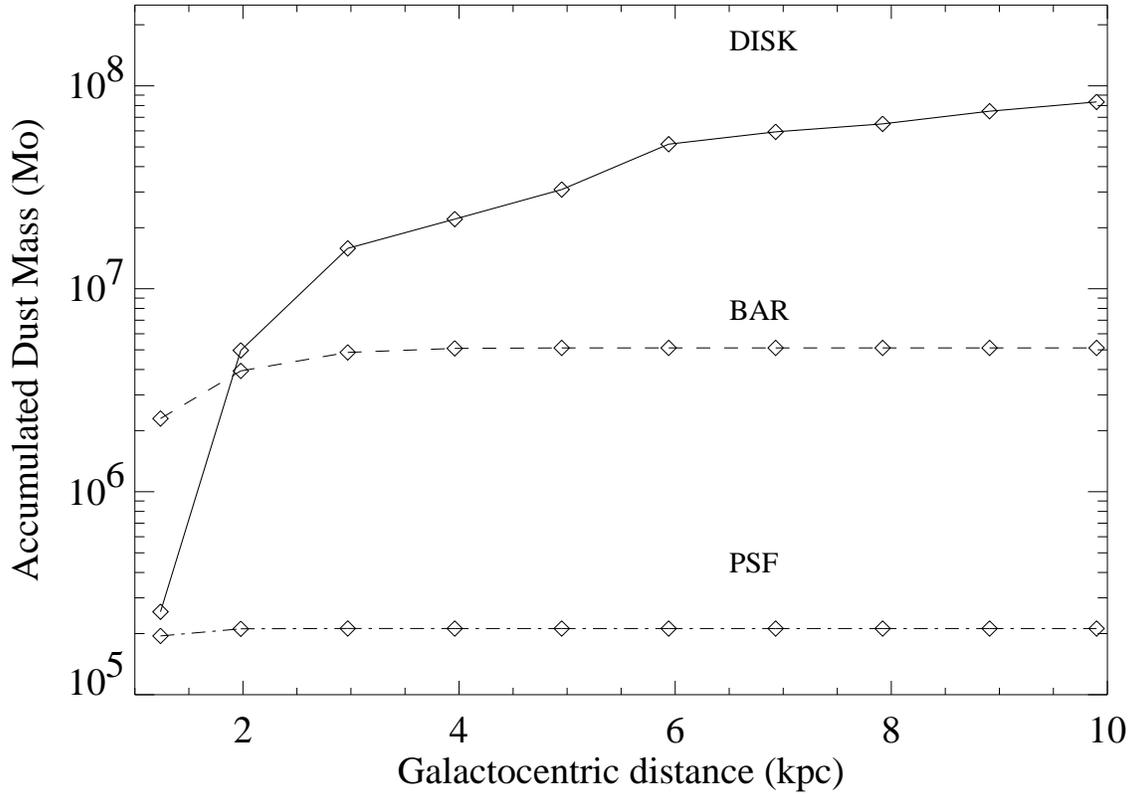}
\caption{Variation  along the semi-major axis of the accumulated mass for 
each morphological component. The disk is represented by a continuous line, 
the bar by a dashed line, and the PSF by a dashed-dotted line. The errors in the
disk and PSF data are of the order of the size of the symbols used. The errors
of the bar are however fairly large ($\sim$1.3$10^7$\Msun) and are not plotted
to avoid making the graph look confusing.}
\label{fi:accumass}
\end{figure}

The computed values for each component are shown in Table \ref{tbl-2}. The total
dust mass of the galaxy is computed as the sum of all contributions. 
The most important contribution 
(about 94\%) to the total mass comes from the very cold dust present in
the disk. The mass estimate depends on the emissivity law adopted; we have
checked that for the emissivity index 1, the resulting mass varies by 
a factor 0.7 to 2 for dust temperatures in the range 18 to 65~K, respectively.
We have also investigated the radial distribution of the dust mass by
integrating over elliptical annuli at a given radius. The cumulative 
distribution per component is plotted in Figure \ref{fi:accumass}. 
It can be noticed that about half of the total dust mass of the galaxy is   
contained in the first 4--5~arcmin. 

\notetoeditor{Figure 8 here.}
The total dust mass value thus estimated
gives a more reliable value than that obtained  from the global SED. 
As an exercise, we have computed the mass using the integrated emission
of the galaxy. For that purpose, the SED have been modeled as 
dust emission from a cold ($T \simeq$ 54 K) and 
a very cold  ($T \simeq$ 20 K) component. In this way, we derive a  
total luminosity of [$3.68~\pm~0.02$]~$\times 10^{10}$~\Lsun~
and total dust 
mass of [$2.92~\pm~0.04$]~$\times 10^{7}$~\Msun,
which are in agreement with the total values obtained for the whole galaxy 
by \citet{Radovich01b} (once the difference in distance has been allowed for).
Note that this value is $\sim 3$ times smaller that the dust mass obtained with the
multiple-component approach.

\begin{figure}
\epsscale{0.8}
\plotone{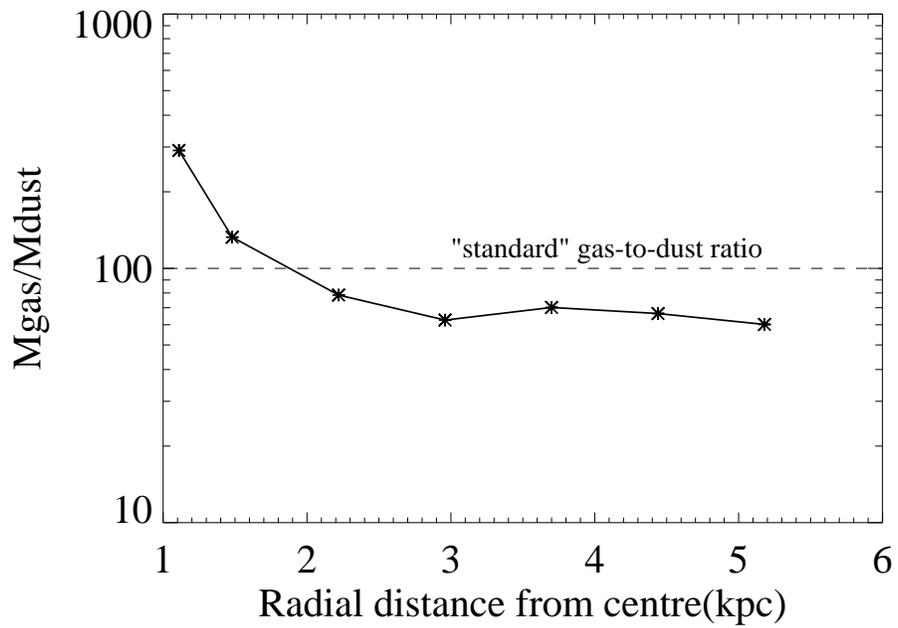}
\caption{Radial variation of gas-to-dust mass ratio. 
For each radius the dust mass is the sum of the dust masses of all 
morphological components (PSF+disk+bar).}
\label{fi:gas2dust}
\end{figure}

We have also computed the gas--to--dust mass ratio for
different galactocentric distances using gas mass data from the literature 
\citep{Scoville85}. 
The result is plotted in Figure \ref{fi:gas2dust}. 
The mass ratio decreases strongly towards the outer part of the galaxy 
approaching the canonical value of our Galaxy. In our Galaxy the gas-to-dust 
mass ratio is $\sim$ 160 \citep{Sodroski94}.
\citet{Devereux90} obtained a value of 1070 for the inner disk (the inner
half of the optical disk, 6.8~kpc) of NGC~253 using {\it IRAS} data. They concluded 
that this value is too high and would indicate that $\sim$ 80--90\% of the 
dust mass in spiral galaxies is radiating at $\lambda$ $>$ 100 $\mu$m. 
At 5.2~kpc from the center, using
our dust mass estimate, we obtain a value of $\sim 60.$
Furthermore, \citet{Houghton97} provided values for molecular and 
atomic hydrogen for $R \le$ 5.6 kpc,  
the total hydrogen mass being $2.4 \times 10^9$~\Msun. We 
find that the global gas-to-dust mass ratio is $\sim 30$,
which is close to the outermost value in Figure \ref{fi:gas2dust}.

\notetoeditor{Figure 10 here.}

Summarizing, we find that the gas--to--dust mass ratio varies greatly along 
the galaxy, from a value of 300 at 1~kpc to $\sim 60$ in 
the outermost part.
The dust mass appears to be distributed towards the outer parts of the galaxy, 
and is rather difficult to detect at $\lambda$$\leqslant$100$\mu$m due to the 
very low temperature. 
The gas to dust mass ratio seems to indicate little excess of dust as 
compared to the canonical value in our Galaxy.

\section{Summary and conclusions}
\label{sec:conclu}

We have used  {\it ISOPHOT} at 180$\mu$m and {\it IRAS} maps to study the spatial distribution of the
far-infrared emission in NGC~253. We have performed an analysis of the radial
brightness profiles in terms of morphological components,
reaching the following conclusions:

\begin{itemize}

\item The radial light profiles show different extensions for the emission
at the different wavelengths. The profiles from 12 to 100~\micron\ show
a central peak, reflecting the nuclear concentration of warm dust due to 
the starburst activity. On the outer parts the profiles correspond to 
an exponential disk. 
At 180~\micron~the slope is rather flat, and corresponds
to exponential disk.
The decay of the light profile at 180~\micron\ is comparable to that seen in the optical 
(B band).
An extended cold dust halo is revealed in the 180\micron\ map. Marginal
evidence is also found at 100\micron.

\item We have identified three main structures: 
an unresolved nuclear component which corresponds to the central starburst, 
a bar, and an exponential disk. 
In addition we have identified a ring at 12 and 25\micron. 

\item The scale length of the disk is about 3.6~kpc and is very similar 
in all the bands studied. 
In all filters (12, 25, 60, and 100 $\mu$m)  we
obtained the same asymptotic limiting values for the ellipticity and the 
position angle: $\epsilon \simeq 0.79 \pm 0.10$ and PA~$\simeq 
52^\circ \pm 7$. These values are in good agreement with those derived from
optical images. We conclude that the scale lengths of both the stars and the
dust are similar.

\item The disk can be characterized by dust emission at 
two temperatures:  a warm
component with dust temperature of $\sim  55$~K and a very cold component 
with temperature of $\sim 16$~K. The first component results dust heated 
by star forming regions in the galaxy disk, while the very cold dust is  
heated by the interstellar radiation field.
This work confirms the presence of a large amount of cold dust in 
the disk of NGC~253, which  contains $\sim 94\%$ of the overall dust  mass.
The estimated total dust mass is $8.2\times 10^7\ M_{\odot}.$

\item We could identify in all the light profiles a nuclear unresolved source,
which was modeled as the PSF of the corresponding filter. 
The SED of this component could be modeled as blackbody emission at $\sim 50$K.
The dust is heated by the intense radiation field produced in the nuclear
starburst. 

\item A far IR bar has also been detected and characterized. This 
bar had been previously detected in the near IR. 
The bar position angle (69.5$^\circ$) 
and its extent are in agreement with the near-IR bar. 
The SED of the bar could  be modeled with 
two components, 
a warm dust component at $\sim$148 K and a cold dust component at $\sim$23 K.

\item In the light profiles at 12 and 25~\micron, we could identify
a ring structure, showing at 3 kpc coincident with the end of the bar. 
The IR color profiles indicate the presence of PAHs at the end of the bar and
coincident with the position of the ring.  
The detection of the ring had been 
previously claimed from both molecular gas (CO) and H$\alpha$ imaging. 

\item The total IR luminosity computed for NGC~253 is 
$4\times 10^{10} {\rm L}_{\odot}$. About half of the total luminosity 
corresponds to the central unresolved component, 
31\% from the disk and 17\% comes from the bar.   

\item  The total star formation rate in NGC~253 is about  5.0~\Msun/yr.
About 70\% comes from the nuclear starburst and the rest
comes from the disk and the bar. 
Estimates of the star formation rates derived from H$\alpha$ photometry
are a factor $\sim 10$ below our estimate, which is consistent with 
previous estimates for optical extinction.  

\item The gas--to--dust 
mass ratio varies appreciably throughout the galaxy from a value
300 in the inner part to $\sim$ 60 in the outermost part. 
Globally NGC~253 seems to be a dust rich galaxy, contrary to what 
was found in previous studies.

\end{itemize}

{\it Acknowledgements.} We thanks an anonymous referee for his very helpful
and detailed report that greatly improved our paper.
This proyect has been partly funded by the spanish DGI (AYA2001-3939-C03-03).
We express our thanks to Antonia Mar\'{\i}a Varela for her comments and 
fruitful discussions. Thanks are due to Prof. John Beckman for 
reading of the text. We also acknowledge the Scientific Editorial Service
of the IAC for corrections.

IPAC is operated by the Jet Propulsion Laboratory (JPL) and California
Institute of Technology (Caltech) for NASA. IPAC is funded by NASA as
part of the IRAS extended mission program under contract to JPL/Caltech.

The COBE datasets were developed by the NASA Goddard Space Flight Center 
under the guidance of the COBE Science Working Group and were provided by 
the NSSDC.

\clearpage

%Figure 1

%Figure 2 : Contours+isophotes over greyscale images

%Figure 3: Light profiles, ellipticity and PA

%Figure 4 : Normalized luminosity distributions

%Figure 5 : Variation of IR colors

%Figure 6

%Figure 7

%Figure 8

%Figure 9

\end{document}